\documentclass{aa} 

\usepackage{textcomp, gensymb}

\usepackage{graphicx}
 \usepackage[varg]{txfonts}

\usepackage[dvipsnames]{xcolor}
\definecolor{VividBlue}{RGB}{0, 163, 224}
\usepackage[colorlinks=true,citecolor=VividBlue,urlcolor=blue,linkcolor=black]{hyperref}
\hypersetup{colorlinks=true, urlcolor=VividBlue}
\usepackage{booktabs}

\usepackage{footmisc}
\usepackage{float}

\makeatletter
\def\@makefnmark{\hbox{\@textsuperscript{\normalfont\color{red}\@thefnmark}}}
\makeatother

\ifdefined\DeclareUnicodeCharacter\DeclareUnicodeCharacter{2212}{-}\fi

\usepackage{tikz}
\usetikzlibrary{shapes.geometric, arrows.meta, positioning, shadows.blur}

\begin{document}

\title{Science-driven design and performance evaluation of a Silicon Laue Component lens for hard X-ray polarimetry}

\author{%
L.~Abalo\inst{\ref{affil:Leiden}} \and
N.~M.~Barri\`ere\inst{\ref{affil:cosine}} \and
M.~W.~Beijersbergen\inst{\ref{affil:Leiden},\ref{affil:cosine}}
}

\institute{
Huygens-Kamerlingh Onnes Laboratory, Leiden University, Postbus 9504, 2300 RA Leiden, The Netherlands \label{affil:Leiden} \email{abalo@physics.leidenuniv.nl}
\and 
cosine Research B.V., Warmonderweg 14, 2171 AH Sassenheim, The Netherlands \label{affil:cosine}
}

\date{Received: -- / accepted: --}

\abstract
{Focusing hard X-rays above $\sim$$80$\,keV remains challenging because the increasingly shallow grazing angles required for reflection limit conventional focusing optics. Although multilayer coatings can extend the useful energy range of grazing-incidence mirrors, at higher energies the required reflection angles become so small that achieving substantial collecting area would demand impractically large mirror diameters and focal lengths, while the achievable reflectivity also decreases. Laue lenses provide an alternative by exploiting crystal diffraction to concentrate hard X-rays onto a compact focal plane.}
{We investigate whether a Laue lens based on Silicon Laue Components (SiLCs), consisting of stacked silicon plates, can provide the collecting area and focal-plane concentration required for polarimetric studies of accreting stellar-mass black holes and strongly magnetized neutron stars in the 80--200\,keV band.}
{We derive reference effective-area requirements from the minimum detectable polarization at 99\% confidence
(${\rm MDP}_{99}$) of an XL-Calibur-inspired focal-plane polarimeter for a 100-mCrab source observed for 100\,ks. We then evaluate an 18-m-focal-length lens combining Si(220) and Si(111) reflections, with a $63''$ diffraction bandpass and a 90\% optical filling factor, using on-axis ray tracing of a perfectly aligned lens with uniform crystal curvatures.}
{An ${\rm MDP}_{99}$ of 5\% requires mean effective areas of approximately 260, 400, and 740\,cm$^{2}$ in the 80--100, 100--130, and 130--200\,keV intervals, respectively. The proposed lens spans radii from 356 to 1453\,mm and provides continuous coverage from 80 to 200\,keV. Its effective area reaches approximately 375\,cm$^{2}$ close to the low-energy boundary and decreases to approximately 200\,cm$^{2}$ at 180\,keV. The focal-plane diameters enclosing 50\% and 90\% of the focused signal are 5.4\,mm ($62''$) and 9.8\,mm ($112''$), respectively. Averaging the simulated effective-area curve over the three reference energy intervals gives approximately 320, 250, and 220\,cm$^{2}$, corresponding to reference ${\rm MDP}_{99}$ values of approximately 4.5\%, 6.4\%, and 9.5\%, respectively.}
{The design reaches the few-percent polarimetric regime below 100\,keV but does not maintain an ${\rm MDP}_{99}$ of 5\% throughout the full 80--200\,keV range under the adopted detector assumptions. It nevertheless remains relevant for substantially more strongly polarized hard components. The results establish the intrinsic on-axis optical performance of the Silicon Laue Component (SiLC) concept and provide a reference configuration for future end-to-end optimization of the optics, detector response, and background.}


\maketitle

\section{Introduction}
\label{section:introduction} 

The hard X-ray band, broadly spanning a few tens to several hundred keV, provides access to some of the most energetic processes in the Universe, including accretion onto compact objects, particle acceleration, and nuclear emission \citep{Vink2012,staubert2019}. Observations in this range remain challenging because the efficiency of conventional focusing optics decreases rapidly with photon energy \citep{Arnaud_2011}. X-ray telescopes such as \textit{Chandra} \citep{Weisskopf_2000}, \textit{XMM-Newton} \citep{Jansen_2001}, and \textit{NuSTAR} \citep{Harrison13longlist} rely on grazing-incidence reflection. Multilayer coatings extend this approach to higher energies, as demonstrated by \textit{NuSTAR}, which focuses X-rays up to 79\,keV, but the progressively smaller grazing angles required at still higher energies make further extension increasingly difficult \citep{Harrison13longlist}. Consequently, observations above this range have relied predominantly on non-focusing instruments, including coded-mask telescopes such as INTEGRAL/IBIS \citep{Ubertini_2003} and \textit{Swift}/BAT \citep{Barthelmy_2005}. By concentrating photons collected over a large aperture onto a small detector region, focusing reduces the detector area contributing background and can therefore provide a substantial gain in sensitivity \citep{Frontera_2010}. Extending this advantage to higher energies requires a different focusing approach.

Laue lenses provide such an approach by using Bragg diffraction in transmission through crystals arranged to concentrate different photon energies onto a common focal plane \citep{Lund_1992,Halloin_2005,Barriere_2009}. Early implementations demonstrated the concept with mosaic crystals \citep{vonBallmoos_2004,Frontera_2008}, while bent crystals subsequently improved diffraction efficiency and radial photon concentration \citep{Frontera_2012,Virgilli_2015,Virgilli_2017}. A practical limitation remains the construction of a complete lens: conventional concepts require large numbers of individual crystals whose diffraction properties must be controlled and whose position and orientation must be set with high accuracy. Silicon Laue components (SiLCs), introduced by \citet{barriere23}, address this challenge by transferring stacking techniques developed for silicon pore optics \citep[SPO;][]{Beijersbergen_2004} to thin monocrystalline silicon plates. A SiLC focuses in both radial and azimuthal directions, allowing the physical dimensions of an element to be largely decoupled from its focal-spot size \citep{barriere23}. Larger diffracting elements can therefore be used, reducing the number of components required to populate a lens while retaining independent control of the focusing geometry and diffraction bandpass.

In this work, we develop and assess a SiLC-based Laue-lens configuration for hard X-ray astronomy. We first discuss the scientific opportunities enabled by sensitive observations in this energy range (Sect.~\ref{section:science_case}) and use polarimetry of compact objects as a representative case from which to derive quantitative requirements on energy coverage, effective area, and focal-plane concentration. We then introduce the SiLC technology (Sect.~\ref{section:silc}) and use these requirements, together with system-level constraints, to define the crystal selection and geometry of the lens (Sect.~\ref{section:requirements_design}). We evaluate its ideal on-axis optical performance through ray-tracing simulations (Sect.~\ref{section:performance}) and compare the resulting effective area with the polarimetric requirements. Finally, we discuss the science reach and limitations of the design and the developments required towards an end-to-end instrument concept (Sect.~\ref{section:summary_outlook}).

\section{Science case}
\label{section:science_case}

The 80--200\,keV band provides access to physical processes that are difficult to probe at lower X-ray energies, particularly in compact objects where Comptonization, non-thermal emission, and strong magnetic fields shape the observed high-energy radiation \citep{nowak11,nobili+08,gotz+06,enoto+17}. In this work, we focus on hard X-ray polarimetry, for which focusing optics can provide a major sensitivity advantage by combining a large collecting area with a compact focal-plane detector \citep{Frontera_2010}. We consider two representative science cases to define the requirements of the lens: Cygnus~X-1, where the 80--200\,keV band probes the transition from the weakly polarized Comptonized spectrum towards an additional high-energy component \citep{mcconnell00,laurent11,Jourdain12b,Chattopadhyay_2024}, and strongly magnetized neutron stars, whose hard emission can be highly polarized and can constrain the geometry and radiation mechanisms of magnetically controlled plasmas \citep{nobili+08,rigoselli2025}.

\subsection{Black-hole X-ray binaries: Cygnus X-1}
\label{subsubsec:cygx1}

In black-hole X-ray binaries, the hard-state spectrum is dominated by Comptonization in a hot plasma, but spectroscopy alone does not uniquely determine the geometry of the emitting region or its connection to the relativistic jet \citep{nowak11}. Polarization provides an independent constraint because different coronal geometries produce different signatures \citep{schnittman10}. In Cygnus~X-1, IXPE measured a polarization degree of $4.01\pm0.20$\% in the 2--8\,keV band, with the polarization angle aligned with the radio jet, favouring a corona extended predominantly in the plane perpendicular to the jet axis \citep{Krawczynski_2022}. XL-Calibur subsequently inferred a polarization degree of $5.0^{+2.7}_{-3.0}$\% over 19--64\,keV in the hard state, consistent with the few-percent polarization measured by IXPE, although the inferred value lies below its 99\% MDP \citep{Awaki_2025}. At higher energies, the thermal Comptonization spectrum turns over around $\sim$$100$\,keV and an additional component is required, commonly associated with either non-thermal Comptonization or synchrotron emission from the jet \citep{mcconnell00,laurent11,Jourdain12b}.

Polarimetry provides a direct way to distinguish these components. Measurements below $\sim$$100$\,keV indicate a low polarization degree consistent with predominantly coronal emission \citep{Chauvin_2018,Krawczynski_2022,Awaki_2025}, whereas INTEGRAL measurements above a few hundred keV reveal a substantially more strongly polarized component \citep{laurent11,Jourdain12b}. AstroSat/CZTI observations indicate that the transition may occur across the intermediate hard-X-ray band: no significant polarization was detected at 100--175\,keV, while polarization degrees of approximately 26\% and 39\% were reported in the 175--230 and 230--380\,keV intervals, respectively, during the hard-intermediate state \citep{Chattopadhyay_2024}. The 80--200\,keV interval therefore covers the weakly polarized regime and the onset of the transition towards the high-energy component, making it particularly valuable for determining how the Comptonized emission connects to the additional hard tail.

\subsection{Strongly magnetized neutron stars}
\label{subsubsec:neutron-stars}

Strongly magnetized neutron stars provide a complementary case in which the magnetic field directly controls the geometry and radiative transfer of the emitting plasma. In accreting X-ray pulsars, matter is channelled towards the magnetic poles, and the observed polarization depends on the magnetic-field geometry, viewing angle, and structure of the accretion column. Energy- and phase-resolved polarization can therefore distinguish between different emission geometries and radiation patterns. X-Calibur observations of GX~301$-$2 provided the first hard-X-ray polarimetric constraints on an accreting neutron star in the 15--35\,keV band, although the limited exposure did not yield a significant detection \citep{Abarr_2020}.

Magnetars provide an especially relevant case at higher energies. Many show persistent hard tails above $\sim$10--20\,keV that extend to several hundred keV \citep{gotz+06,enoto+17}, whose origin remains uncertain. Proposed mechanisms include bremsstrahlung from surface layers heated by returning currents, synchrotron emission from electron--positron pairs, and resonant Compton up-scattering in a twisted magnetosphere \citep{nobili+08}; these scenarios predict different polarization signatures. Simultaneous IXPE and \textit{NuSTAR} observations of 1E~1841$-$045 showed that a spectropolarimetric decomposition of the broadband emission requires a highly polarized hard power-law component, with a polarization degree exceeding 65\%, favouring a synchrotron/curvature-radiation interpretation \citep{rigoselli2025}. This component extends beyond the IXPE band and remains unexplored polarimetrically at the energies where it dominates. Extending energy-resolved polarimetry to 80--200\,keV would test whether the high polarization inferred for the hard component persists through the hard X-ray tail.

\section{Silicon Laue Components}
\label{section:silc}

Silicon Laue Components (SiLCs) are self-supporting stacks of thin monocrystalline silicon plates that combine Bragg diffraction with two-dimensional focusing \citep{barriere23}. The plates are produced from double-sided super-polished silicon wafers and assembled by direct silicon bonding, without adhesives, while conforming to the geometry of a mandrel. Once bonded, the elastically deformed plates retain the imposed geometry. The manufacturing approach is inherited from SPO, developed for lightweight X-ray optics using semiconductor-grade silicon wafers \citep{barriere22}. It therefore combines high-quality crystalline material with manufacturing and stacking processes developed for serial production.

The optical properties of a SiLC are controlled by three geometrical parameters. A wedge between consecutive plates progressively changes the orientation of their crystallographic planes so that the diffracted beams from the stacked plates are confocal in the radial direction. Sagittal curvature provides focusing in the orthogonal, azimuthal direction. The physical dimensions of a SiLC can therefore be substantially larger than its focal spot. Finally, meridional curvature changes the orientation of the diffracting planes continuously along each plate and thus sets its angular diffraction bandpass. The wedge, sagittal curvature, and meridional curvature consequently control the confocality, two-dimensional focusing, and energy acceptance of the component, respectively \citep{barriere23}.

\section{Optical requirements and lens design}
\label{section:requirements_design}

\subsection{Design requirements}
\label{subsec:design_requirements}

The science cases discussed in Sect.~\ref{section:science_case} motivate three main requirements for the optical design. First, we require continuous coverage over 80--200\,keV. For Cygnus~X-1, this interval covers the weakly polarized hard-state emission and the onset of the transition towards the more strongly polarized high-energy component, while for magnetars it probes the hard non-thermal emission beyond the energy range currently accessible to focusing polarimeters. Second, the effective collecting area must provide sufficient photon statistics for meaningful polarimetric measurements. Finally, the collected photons should be concentrated onto a compact focal-plane region, thereby limiting the detector volume and instrumental background associated with the source extraction region. For the non-imaging scattering polarimeter considered here, the relevant optical requirement is the physical encircled-energy diameter at the focal plane rather than angular imaging resolution. Since the targets are known point sources observed in pointed mode, the design is optimized for on-axis collecting area and focal-plane concentration. 

We quantify the collecting-area requirement using the MDP at 99\% confidence,
\begin{equation}
{\rm MDP}_{99} =
\frac{4.29}{\mu R_{\rm src}}
\sqrt{\frac{R_{\rm src}+R_{\rm bkg}}{t_{\rm obs}}},
\end{equation}
where $\mu$ is the modulation factor, $R_{\rm src}$ and $R_{\rm bkg}$ are the source and background count rates, and $t_{\rm obs}$ is the exposure time. As a reference case, we adopt a 100-mCrab source with a Crab-like spectral shape. For one Crab, we use the low-energy branch of the spectral model of \citet{Dirson_2023},
\begin{equation}
\frac{{\rm d}N}{{\rm d}E} = N_0
\left(\frac{E}{100\,{\rm keV}}\right)^{-2.095}
\exp\!\left(-\frac{E}{1360\,{\rm keV}}\right),
\end{equation}
with $N_0=6.40\times10^{-4}$\,ph\,cm$^{-2}$\,s$^{-1}$\,keV$^{-1}$, and scale the flux to 0.1 Crab for the reference source. Here $N_0$ is the model coefficient; the photon flux density at 100\,keV also includes the exponential factor. The band-integrated photon fluxes of the reference source are $1.515\times10^{-3}$, $1.343\times10^{-3}$, and $1.467\times10^{-3}$\,ph\,cm$^{-2}$\,s$^{-1}$ in the three energy intervals, respectively. 

For an XL-Calibur-inspired configuration \citep{Abarr_2021} with an 8-cm-long Be scattering rod of 1\,cm diameter and 2-mm-thick CZT absorbers, the mean detection efficiencies are 0.77, 0.57, and 0.28 in the 80--100, 100--130, and 130--200\,keV bands, respectively. The modulation factor $\mu=0.5$ is considered constant across the energy bands, and a background rate of 0.01\,counts\,s$^{-1}$ is used (Kun Hu, private communication). With an exposure of 100\,ks, these assumptions lead to mean effective-area requirements of approximately 260, 400, and 740\,cm$^{2}$ for an ${\rm MDP}_{99}$ of 5\%. Reaching an ${\rm MDP}_{99}$ of 1\% would instead require approximately 6300, 9600, and 18000\,cm$^{2}$, respectively.

For these reference estimates, $R_{\rm src}=\bar\epsilon\,\bar A_{\rm eff}\int_{E_1}^{E_2}({\rm d}N/{\rm d}E)\,{\rm d}E$, treating the detection efficiency and effective area as constant within each band and taking $R_{\rm bkg}=0.01$\,counts\,s$^{-1}$ in each band. The source count rates required for 5\% and 1\% MDP are 0.304 and 7.37\,counts\,s$^{-1}$, respectively. These values should be regarded as reference optical requirements rather than as an end-to-end instrument sensitivity. The detector geometry adopted above was optimized for measurements in the 10--80\,keV range; a focal-plane polarimeter designed specifically for 80--200\,keV could use thicker scattering and absorbing elements and would have a different efficiency and background response. The calculation nevertheless provides a transparent benchmark against which the optical collecting area of the lens can be assessed.

The focal length is selected by combining the optical trade-off with the system envelope. At fixed focal-plane extent, longer focal lengths require narrower diffraction bandpasses but provide a larger geometric collecting area and, in the trade study, a larger effective area for the same EE90. We consequently favour the longest focal length compatible with the adopted telescope dimensions. For a fixed photon energy and crystallographic reflection, the required lens radius scales linearly with focal length. The system constraints impose a focal length below 20\,m and a lens radius below 1.5\,m. At $f=18$\,m, the outermost required radius is 1.453\,m. Scaling this radius to a focal length of 20\,m gives approximately 1.61\,m, corresponding to a lens diameter of about 3.23\,m and therefore exceeding the adopted 3-m envelope. We consequently adopt 18\,m as a near-maximum focal length compatible with the system-level constraints. 

\subsection{Lens geometry and crystal selection}
\label{subsec:lens_geometry}

\begin{figure}
    \centering
    \includegraphics[width=\columnwidth]{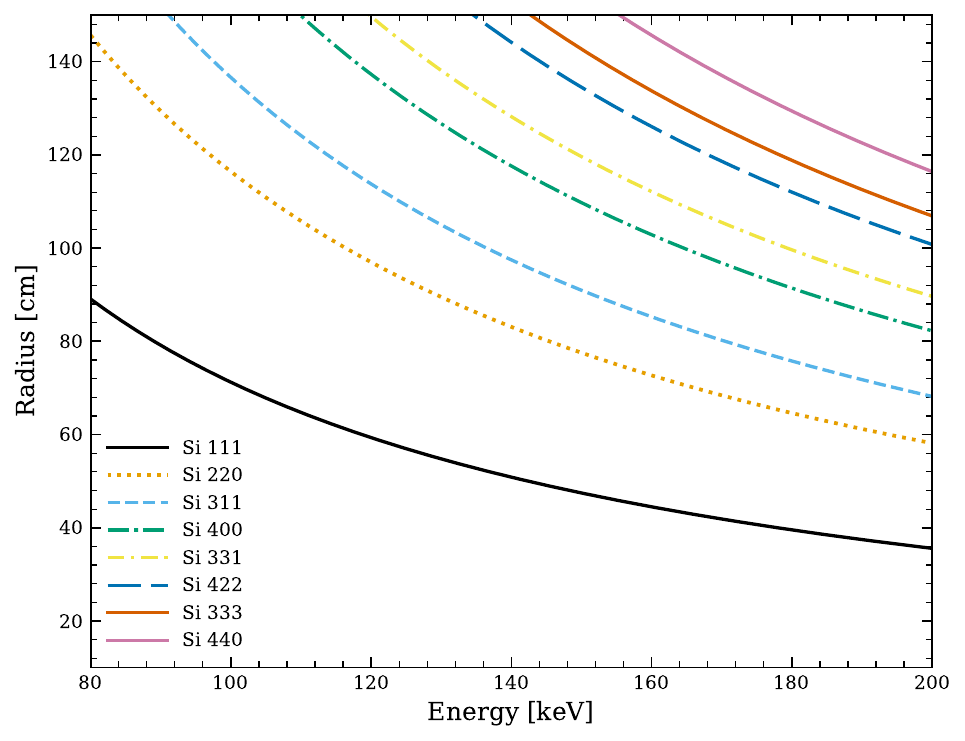}
    \caption{Radial position required to focus an on-axis source at a focal length of 18\,m as a function of photon energy for the principal crystallographic reflections of silicon. The adopted design uses Si(220) over 80--200\,keV and adds Si(111) at smaller radii. }
    \label{fig:radius_energy}
\end{figure}

In a Laue lens, the photon energy that can be diffracted by a given crystal depends on the spacing of the selected crystallographic planes and on the Bragg angle. For an on-axis source at infinity, the Bragg angle also determines the radial position at which the crystal must be placed for the diffracted photons to converge at the focal point. The photon energy, crystallographic reflection, focal length, and radial position are therefore directly coupled. Figure~\ref{fig:radius_energy} shows this relation for the principal silicon reflections at the adopted focal length of 18\,m.

We combine Si(220) and Si(111) to provide continuous coverage over 80--200\,keV while mitigating the decline in effective area towards high energies. Si(220) forms the baseline reflection across the full energy interval: its larger Bragg angle places the corresponding elements at larger radii, providing a larger geometric collecting area. Si(111), which offers higher diffraction efficiency but smaller Bragg angles, is added at smaller radii from $\sim$125\,keV upwards to increase the high-energy response. In addition, Si(220) components contribute through the allowed second-order Si(440) reflection above $\sim$160\,keV, whereas the corresponding Si(222) reflection is forbidden. The resulting lens occupies radii from 356 to 1453\,mm.

The full-lens design uses 0.775-mm-thick silicon plates, corresponding to the standard thickness of 300-mm wafers used by the SiLC manufacturing process. The diffracting ribs have a nominal transverse width of 0.55\,mm. The crystal path length is not fixed across the lens: it is optimized as a function of energy and diffraction bandpass to balance diffraction efficiency against absorption, ranging from approximately 13 to 22\,mm for Si(111) and from 8 to 23\,mm for Si(220).

\subsection{Diffraction bandpass}
\label{subsec:bandpass}

The meridional curvature determines the angular diffraction bandpass of each SiLC and introduces the principal trade-off between collecting area and focal-plane concentration. Increasing the bandpass broadens the range of Bragg angles sampled within each plate and increases the integrated reflectivity, but it simultaneously broadens the focal-plane distribution \citep{barriere23}. Therefore, the bandpass must be chosen jointly with the focal length and the maximum focal-plane footprint rather than optimized for effective area alone.

We adopt a bandpass of $63''$. This value follows from the focal-length--bandpass trade study under the requirement that 90\% of the focused signal remain within a diameter of approximately 10\,mm, corresponding to the scale of the scattering element of an XL-Calibur-like polarimeter. For a fixed EE90, shorter focal lengths permit broader bandpasses, whereas longer focal lengths require narrower bandpasses; despite the smaller bandpass, the larger geometric aperture obtained at longer focal lengths results in a larger effective area. The adopted combination of $f=18$\,m and $\Omega=63''$ is the baseline configuration resulting from this trade-off while remaining compatible with the focal-plane and telescope-size constraints. 

\section{Optical performance}
\label{section:performance}

We evaluate the optical performance of the lens baseline configuration defined in Sect.~\ref{section:requirements_design} using ray tracing that incorporates the dynamical theory of diffraction for curved diffracting-plane crystals. The lens is represented by concentric rings of Si(220) and Si(111) elements, including their energy-dependent diffraction efficiency and absorption in silicon. We assume that 90\% of the geometric annular area between radii of 356 and 1453\,mm is populated by active SiLC material. The calculations describe an ideal optical assembly: the crystal curvatures are assumed to be uniform, and no structural deformation, SiLC misalignment, wedge error, stacking error, or other manufacturing tolerance is included. The results represent the intrinsic optical performance of the baseline design rather than the expected performance of an as-built telescope.

We restrict the present analysis to an on-axis source at infinity. The compact-object applications considered here are pointed observations of sources with accurately known positions and do not require survey imaging over a wide field of view. The off-axis response of Laue lenses has been investigated previously and leads to a less concentrated focal-plane response, reduced detection significance, and a degeneracy in source position that can be mitigated through dedicated pointing strategies \citep[e.g.][]{Barriere_2009}. We therefore focus here on the on-axis response relevant to the targeted polarimetric observations; optimization of off-axis imaging performance is beyond the scope of the present work.

\begin{figure}
    \centering
    \includegraphics[width=\columnwidth]{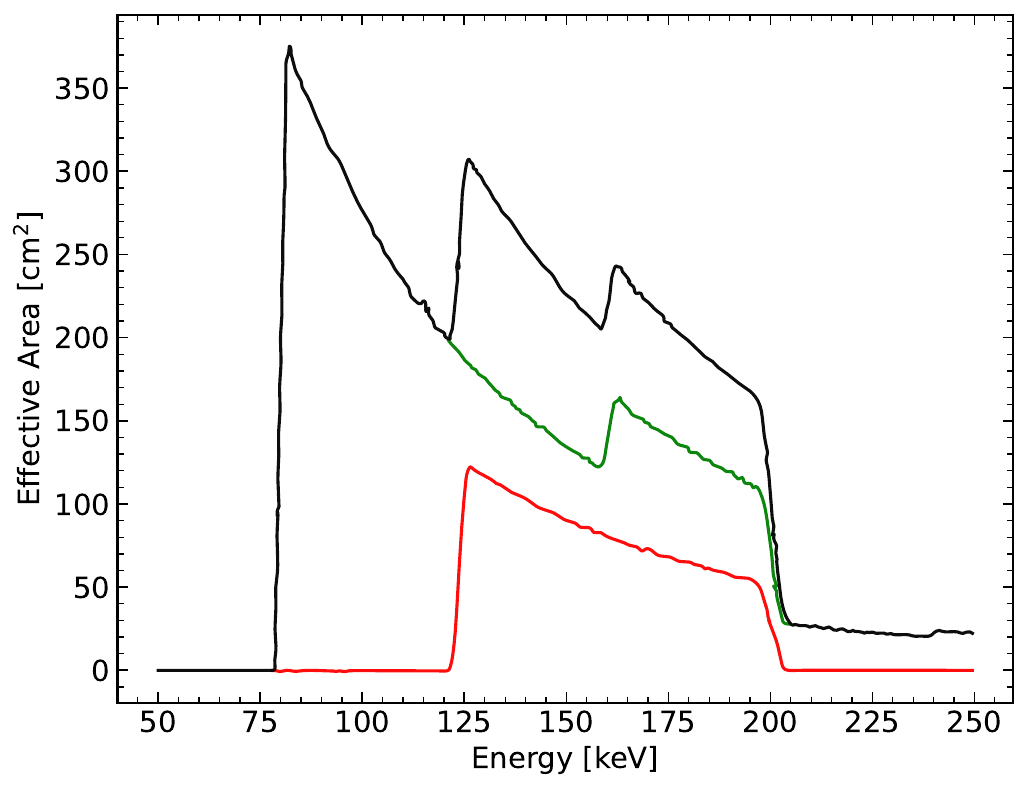}
    \caption{Ideal on-axis effective area of the SiLC Laue-lens design as a function of photon energy, assuming a 90\% optical filling factor. The Si(220) (green) and Si(111) (red) contributions are shown separately together with their total (black).}
    \label{fig:effective_area}
\end{figure}

Figure~\ref{fig:effective_area} shows the resulting on-axis effective area. The total effective area reaches approximately 375\,cm$^{2}$ close to the low-energy boundary, is about 325\,cm$^{2}$ at 90\,keV, and decreases to approximately 200\,cm$^{2}$ at 180\,keV. The overall decline with energy reflects the combination of the decreasing geometric area available at the corresponding crystal radii and the energy dependence of the diffraction efficiency.

The simulated effective area can be compared directly with the polarimetric benchmark derived in Sect.~\ref{subsec:design_requirements}. Averaging the simulated effective-area curve over the same three energy intervals gives approximately 320, 250, and 220\,cm$^{2}$ in the 80--100, 100--130, and 130--200\,keV bands, respectively. The first value exceeds the 260\,cm$^{2}$ required for an MDP of 5\% in the reference 100-mCrab, 100-ks observation, whereas the two higher-energy values fall below the corresponding requirements of 400 and 740\,cm$^{2}$. Using the same MDP relation and treating the quoted mean effective area as constant within each band gives approximate ${\rm MDP}_{99}$ values of 4.5\%, 6.4\%, and 9.5\%, respectively. The baseline lens therefore reaches the targeted few-percent polarimetric sensitivity below 100\,keV but does not maintain a 5\% MDP throughout the full 80--200\,keV band.

The ideal focal-plane distribution also satisfies the physical concentration requirement imposed by the polarimeter. The diameter enclosing 50\% of the focused signal is 5.4\,mm, corresponding to a half-energy width of approximately $62''$ at a focal length of 18\,m, while 90\% of the signal is enclosed within a diameter of 9.8\,mm, or approximately $112''$. The latter remains below the adopted EE90 requirement of 10\,mm. The reference MDP estimates above do not apply an additional focal-plane interception factor; an end-to-end calculation must convolve the energy-dependent effective area, focal distribution, and detector response. For the non-imaging polarimeter considered here, this physical focal-plane concentration is more relevant than angular imaging resolution because it determines the fraction of source photons intercepted by the scattering element and the detector volume contributing to the background.

These results clarify which aspects of the representative science cases in Sect.~\ref{section:science_case} can be addressed by the baseline design. For Cygnus~X-1, the 80--100\,keV interval reaches a reference MDP of approximately 4.5\% and can therefore probe whether the weak polarization observed at lower energies persists towards the thermal-Comptonization turnover. At higher energies, the reference MDP increases to approximately 6.4\% in the 100--130\,keV band and 9.5\% in the 130--200\,keV band. These values do not provide a uniform 5\% sensitivity throughout the full energy range, but they remain below the tens-of-percent polarization fractions reported as the additional high-energy component emerges. The magnetar case offers an even larger polarization contrast: the hard power-law component inferred for 1E~1841$-$045 has a polarization degree exceeding 65\%. Whether such polarization can be measured for a particular magnetar will also depend on its hard-X-ray spectrum and flux and therefore requires a source-specific end-to-end detector calculation. 

\section{Summary and outlook}
\label{section:summary_outlook}

In this work, we used hard X-ray polarimetry of compact objects as the science case from which to derive quantitative requirements for a SiLC-based Laue lens focusing between 80 and 200\,keV. Adopting an XL-Calibur-inspired detector configuration, we derived the mean effective areas required to reach an ${\rm MDP}_{99}$ of 5\% in three energy intervals (80--100, 100--130, and 130--200\,keV) for a 100-mCrab source with a Crab-like spectrum observed for 100\,ks. Within the constraint of a lens diameter below 3\,m, the design explored here achieves this threshold only in the lowest energy interval, where the reference ${\rm MDP}_{99}$ is approximately 4.5\%. The available effective area corresponds to ${\rm MDP}_{99}$ values of approximately 6.4\% and 9.5\% in the medium- and high-energy intervals, respectively. The achieved collecting area nevertheless remains relevant for the substantially more strongly polarized components discussed in Sect.~\ref{section:science_case}.

These results illustrate the demanding photon statistics of hard X-ray polarimetry: under the adopted source, exposure, and detector assumptions, reaching an ${\rm MDP}_{99}$ of 5\% above 100\,keV remains difficult even for the ideal optical assembly considered here, with a lens diameter of approximately 2.9\,m and a focal length of 18\,m. This limitation reflects the trade-off between collecting area and focal-plane concentration within the imposed system envelope: increasing the SiLC bandpass increases the integrated diffraction response and hence the effective area, but broadens the focal-plane distribution, while increasing the focal length also increases the physical size of the lens. A polarimeter accepting a wider focused beam could therefore be investigated as part of a joint optimization of the optics and detector, accounting for changes in detection efficiency, modulation factor, and instrumental background. Such an optimization is beyond the scope of the present study.

The SiLC concept is not intrinsically restricted to 200\,keV. Extending the operating range would require a new optimization of the crystal material, crystallographic reflections, radial allocation, and focal-plane response, and materials such as germanium could be considered for this purpose. Hybrid configurations could therefore extend the focusing response towards the soft gamma-ray regime and enable additional continuum and nuclear-line applications.

The performance reported here represents an ideal optical assembly: SiLC curvature and wedge errors, component misalignment, structural deformation, thermo-mechanical effects, and pointing uncertainties may broaden the focal distribution or reduce the effective area. Incorporating these effects, together with an optimized focal-plane response and instrumental background, is the next step towards an end-to-end telescope concept. In the meantime, the 80--200\,keV configuration presented here should be regarded as a quantitatively defined reference design for SiLC-based hard X-ray polarimetry rather than as a complete mission architecture.

\begin{acknowledgements}
This work was partially conducted under ESA Technology Development Element contract 4000144487/24/NL/AR/ahh ``OPTICAL ELEMENT TO FOCUS SOFT GAMMA RAYS'' (ASTROLAUE1). The authors thank Kun Hu and Henric Krawczynski at Washington University in St.~Louis for the information regarding the XL-Calibur parameters and performances.
\end{acknowledgements}

\bibliographystyle{aa}
\bibliography{references}

\end{document}